\documentstyle[aps,psfig,multicol,prb]{revtex}
\newcommand{\bcols}{\ifpreprintsty\else\begin{multicols}{2}\fi}
\newcommand{\ecols}{\ifpreprintsty\else\end{multicols}\fi}
\begin{document}
\draft
\title{Effect of inelastic scattering on the average Coulomb-blockade peak
height in quantum dots}
\author{C. W. J. Beenakker$^{1}$, H. Schomerus$^{1}$, and P. G.
Silvestrov$^{1,2}$}
\address{$^1$ Instituut-Lorentz, Universiteit Leiden,
P.O. Box 9506, 2300 RA Leiden, The Netherlands\\
$^2$ Budker Institute of Nuclear Physics, 630090 Novosibirsk, Russia
}
\date{25 October 2000}
\maketitle
\begin{abstract}
The average height of the Coulomb-blockade conductance peaks for chaotic
elastic scattering is known to increase by a factor of 4/3 upon breaking
time-reversal symmetry. We calculate the temperature dependence of this factor
in the regime that the inelastic scattering rate $\Gamma_{\rm in}$ is greater
than the mean tunneling rate $\Gamma_{\rm el}$, which itself is less than the
mean level spacing $\Delta$. Comparison with recent experimental data by Folk
{\em et al.\/} (cond-mat/0008052) demonstrates that $\Gamma_{\rm in}$ lies
below $\Gamma_{\rm el}$ and hence also below $\Delta$, consistent with the
low-energy suppression of inelastic electron-electron scattering in quantum
dots.
\end{abstract}
\pacs{PACS numbers: 73.23.Hk, 73.23.-b, 73.50.Bk, 73.50.Gr}
\bcols

Inelastic electron-electron scattering in a quantum dot broadens the
single-particle excitation levels by an amount $\hbar\Gamma_{\rm in}$. This
broadening vanishes at low excitation energies $\varepsilon$ and remains less
than the mean level spacing $\Delta$ as long as $\varepsilon$ is below the
Thouless energy.\cite{Siv94a,Alt97} Early Coulomb-blockade experiments by Sivan
{\em et al.}\cite{Siv94b} agreed with this theoretical prediction, but recent
experiments by Folk {\em et al.}\cite{Fol00} were interpreted as being
inconsistent with it.

Inelastic scattering can be detected by the broadening of the single-particle
density of states, as was done in Ref.\ \onlinecite{Siv94b}. Ref.\
\onlinecite{Fol00}, instead, used the temperature dependence of the height of
the Coulomb-blockade peaks in the conductance. For fully elastic and chaotic
scattering the average height is increased by a temperature-independent factor
of 4/3 upon application of a magnetic field.\cite{Jal92,Alh98} Folk {\em et
al.\/} measured a suppression of this enhancement factor when the thermal
energy $kT$ became larger than $\Delta$. They concluded from this strong
temperature dependence that the dephasing rate\cite{note1} in quantum dots is
larger than $\Delta/\hbar$ at excitation energies well below the Thouless
energy, in apparent contradiction with the theoretical expectation. However, in
the absence of a quantitative prediction for the temperature dependence of the
Coulomb-blockade peak height, it is difficult to decide whether the observed
temperature dependence is actually stronger than expected.

What we will do here is use the semiclassical theory of the Coulomb
blockade\cite{Bee91} to obtain the temperature dependence in the regime
$\Gamma_{\rm el}\ll\Gamma_{\rm in}$, with $\Gamma_{\rm el}$ the mean
(elastic) tunnel rate into the quantum dot. We call this the regime of
strong inelastic scattering, where ``strong'' means strong enough to
thermalize the distribution of the electrons among the levels in the
quantum dot. Both $\Gamma_{\rm el}$ and $\Gamma_{\rm in}$ should be
less than $kT$, so that we are allowed to use rate equations based
on sequential tunneling. The condition for the Coulomb blockade
is $\Gamma_{\rm el}\ll\Delta/\hbar$ and $kT\ll e^{2}/C$, with $C$
the capacitance of the quantum dot. We find that the experimental
temperature dependence\cite{Fol00} is actually much {\em weaker\/}
than predicted by the theory for strong inelastic scattering. Therefore,
$\Gamma_{\rm in}\lesssim\Gamma_{\rm el}\ll\Delta/\hbar$ and there is no
disagreement between the experimental data of Ref.\ \onlinecite{Fol00}
and the theoretical expectation of a low-energy suppression of inelastic
electron-electron scattering in quantum dots.\cite{Bro00}

Starting point of our analysis is a pair of expressions from Ref.\
\onlinecite{Bee91} for the $N$-th conductance peak in the two cases of purely
elastic scattering ($G_{\rm el}$) and strong inelastic scattering ($G_{\rm
in}$):
\begin{eqnarray}
G_{\rm el}&=&\frac{e^{2}}{kT}P_{\rm eq}(N) \left\langle\frac{\Gamma^{\rm
l}\Gamma^{\rm r}} {\Gamma^{\rm l}+\Gamma^{\rm
r}}\right\rangle_{N},\label{Gelformula}\\
G_{\rm in}&=&\frac{e^{2}}{kT}P_{\rm eq}(N) \frac{\langle\Gamma^{\rm
l}\rangle_{N}\langle\Gamma^{\rm r}\rangle_{N}} {\langle\Gamma^{\rm
l}+\Gamma^{\rm r}\rangle_{N}}.\label{Ginformula}
\end{eqnarray}
The spectral average of the elastic tunnel rate $\Gamma_{p}^{\rm l,r}$ into the
left or right reservoir is defined by
\begin{equation}
\langle\Gamma^{\rm l,r}\rangle_{N}=\sum_{p}\Gamma_{p}^{\rm l,r}[1-F_{\rm
eq}(E_{p}|N)]f(E_{p}-\mu).\label{averagedef}
\end{equation}
The equilibrium distributions $P_{\rm eq}(N)$ and $F_{\rm eq}(E_{p}|N)$ give,
respectively, the {\em a priori\/} probability to find $N$ electrons in the
quantum dot and the conditional probability to find level $p$ occupied by one
of the $N$ electrons. (These functions are obtained from the Gibbs distribution
in the canonical ensemble.) The function $f(E_{p}-\mu)$ is the Fermi-Dirac
distribution, with $\mu$ an externally tunable parameter that depends linearly
on the gate voltage.

If $\Gamma_{\rm in}\ll\Gamma_{\rm el}$ one may neglect inelastic scattering and
use Eq.\ (\ref{Gelformula}), while if $\Gamma_{\rm el}\ll\Gamma_{\rm in}$ one
should use Eq.\ (\ref{Ginformula}). The key difference between the two
equations is that for $G_{\rm el}$ the fraction $\Gamma^{\rm l}_{p}\Gamma^{\rm
r}_{p}/(\Gamma^{\rm l}_{p}+\Gamma^{\rm r})$ as a whole is averaged over the
spectrum, while for $G_{\rm in}$ the numerator and denominator are averaged
separately. Since the spectral average extends over about $kT/\Delta$ levels,
the difference between $G_{\rm el}$ and $G_{\rm in}$ vanishes if $kT$ becomes
less than $\Delta$.

In a chaotic quantum dot, the tunnel rates $\Gamma^{\rm l}_{p}$ and
$\Gamma^{\rm r}_{q}$ fluctuate independently according to the Porter-Thomas
distribution $P(\Gamma)\propto\Gamma^{\beta/2-1}\exp(-\beta\Gamma/2\Gamma_{\rm
el})$. (We assume tunneling through two equivalent single-channel point
contacts, with energy-independent mean tunnel rate $\Gamma_{\rm el}$.) The
index $\beta=1$ (2) in the presence (absence) of a time-reversal-symmetry
breaking magnetic field. The mean height $\overline{G^{\rm max}_{\rm el}}$ of
the Coulomb-blockade peak for elastic scattering increases upon breaking
time-reversal symmetry, by a temperature-independent factor of
4/3.\cite{Jal92,Alh98} Inelastic scattering introduces a temperature
dependence, which we can study using Eq.\ (\ref{Ginformula}).

Qualitatively, the effect of inelastic scattering on the 4/3-enhancement factor
can be understood as follows. The spectral average $\langle\cdots\rangle_{N}$,
defined precisely in Eq.\ (\ref{averagedef}), can be approximated by an average
over $kT/\Delta$ levels around the Fermi energy in the quantum dot containing
$N$ electrons. If $kT\gg\Delta$ the spectral average becomes equivalent to an
ensemble average. The ensemble averages of $\Gamma_{p}^{\rm l}$ and
$\Gamma_{p}^{\rm r}$ are both equal to the $\beta$-independent value
$\Gamma_{\rm el}$, so the peak height (\ref{Ginformula}) for strong inelastic
scattering simplifies to $G_{\rm in}\approx \frac{1}{2}\Gamma_{\rm
el}(e^{2}/kT)P_{\rm eq}(N)$ --- independent of whether time-reversal symmetry
is broken or not. This explains why the enhancement factor drops from 4/3 to 1
as $kT$ becomes larger than $\Delta$ in the case of strong inelastic
scattering.

\begin{figure}
\centerline{\psfig{figure=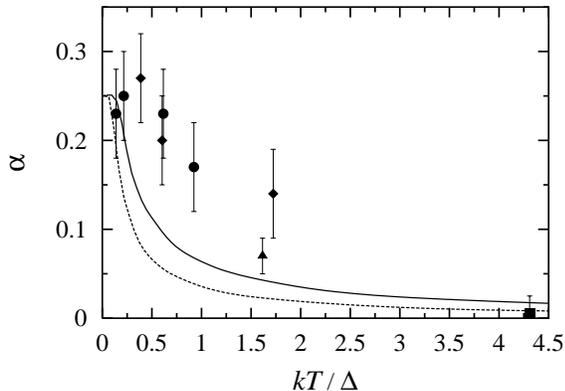,width=8cm}\medskip}
\caption[]
{Temperature dependence of the parameter $\alpha$ defined in Eq.\
(\protect\ref{alphadef}). The curves are calculated from Eq.\
(\protect\ref{Ginformula}), either for spin-degenerate levels (solid) or for
non-degenerate levels (dashed). The markers with error bars are experimental
data for GaAs quantum dots from Folk {\em et al.}\protect\cite{Fol00} The area
of the dot is 0.25, 0.7, 3, and 8~$\mu{\rm m}^{2}$ for, respectively, circle,
diamond, triangle, and square markers.
\label{folkfigure}
}
\end{figure}

For a quantitative comparison, we have plotted in Fig.\ 1 the temperature
dependence of the parameter
\begin{equation}
\alpha=1-\overline{G^{\rm max}_{\rm in}}(\beta=1)/\,\overline{G^{\rm max}_{\rm
in}}(\beta=2),\label{alphadef}
\end{equation}
which drops from 1/4 to 0 as $kT$ becomes larger than $\Delta$. The solid curve
is for equally-spaced spin-degenerate levels ($E_{2p}=E_{2p-1}=p\Delta$,
$\Gamma_{2p}=\Gamma_{2p-1}$). Because the spin degeneracy might be lifted
spontaneously,\cite{Bla97} we also show for comparison the case of
equally-spaced non-degenerate levels ($E_{p}=p\Delta/2$, all $\Gamma_{p}$'s
independent). In either case $\Delta$ is defined as the mean level spacing of a
single spin degree of freedom. We see that the temperature dependence is
stronger for non-degenerate levels. An even stronger temperature dependence
(not shown) is found if, instead of equally spaced levels, we would use a
Wigner-Dyson distribution. The data points are the experimental results of Folk
{\em et al.,}\cite{Fol00} for GaAs quantum dots of four different areas. The
values of $\Delta$ used are those given in Ref.\ \onlinecite{Fol00}, estimated
from the area $A$ and the two-dimensional density of states
($\Delta=2\pi\hbar^{2}/mA$, with $m$ the effective mass of the electrons).
There is therefore no adjustable parameter in the comparison between theory and
experiment.

It is clear from Fig.\ 1 that the experimental temperature dependence is much
weaker than the theoretical prediction, regardless of whether we include spin
degeneracy or not. We have found that the theory would fit the data within the
errorbars if we would rescale $kT/\Delta$ by a factor of 3 (with spin
degeneracy) or a factor of 5 (without spin degeneracy). Such a large factor is
beyond the experimental uncertainty in level spacing or temperature. We
conclude that the inelastic scattering rate is well below $\Gamma_{\rm el}$ and
$\Delta/\hbar$ for a range of energies within $kT$. One possible explanation of
the deviation of our theoretical curves from the experimental data would be
that only the high-lying levels have equilibrated, while the low-lying levels
have not. Such an explanation would be consistent with the scenario put forward
in Ref.\ \onlinecite{Alt97}, according to which the discreteness of the
spectrum prevents the low-lying levels to equilibrate on an arbitrarily long
time scale.

We conclude with two suggestions for future research on this topic. From the
theoretical side, it would be useful to generalize Ref.\ \onlinecite{Bee91} to
arbitrary ratio of $\Gamma_{\rm el}$ and $\Gamma_{\rm in}$ [going beyond the
two limits of large and small $\Gamma_{\rm el}/\Gamma_{\rm in}$ given in Eqs.\
(\ref{Gelformula}) and (\ref{Ginformula})]. From the experimental side, it
would be of interest to compare data for the temperature dependence of $\alpha$
for different values of $\Gamma_{\rm el}$, that is to say, for different
heights of the tunnel barriers separating the quantum dot from the electron
reservoirs. We would expect the data points in Fig.\ 1 to approach the
theoretical curves as the tunnel barriers are made higher and higher, giving
more precise information on the rate of inelastic scattering.

This work was supported by the Dutch Science Foundation NWO/FOM. We have
benefitted from correspondence with P. W. Brouwer and C. M. Marcus.

\ecols
\end{document}